\documentclass[aps,prb,twocolumn,eqsecnum,epsfig]{revtex4}

\usepackage{graphicx}
\begin{document} 
\title{Charge Screening Effect in Metallic Carbon Nanotubes}
\author{K. Sasaki}
\affiliation{Department of Physics, Tohoku University, Sendai 980-8578, Japan}
%%%%%%%%%%%%%%%%%%%%%%%%%%

\begin{abstract}
Charge screening effect in metallic carbon nanotubes is investigated in a model including 
the one-dimensional long-range Coulomb interaction. It is pointed out that an external
charge which is being fixed spatially is screened by internal electrons so that 
the resulting object becomes electrically neutral. We found that the screening length is
given by about the diameter of a nanotube.
\end{abstract}
\maketitle
%\pacs{}
%%%%%%%%%%%%%%%%%%%%%%
\section{introduction}

Recently carbon nanotubes(CNTs)~\cite{Iijima}
have attracted much attention from various points of view.
Especially their unique mechanical, electrical, chemical properties have
stimulated many people's interest in the analysis of CNTs~\cite{SDD,Dekker}. 
They have exceptional strength and stability,
and they can exhibit either metallic or semiconducting properties
depending on the diameter and helicity~\cite{MDW,Wildoer}.
Because of their small size, properties of CNTs should be 
governed by the law of quantum mechanics.
Therefore it is quite important to understand the 
quantum behavior of electrons in CNTs.

However in low energy region( $< 10^4 [\rm{K}] $), we do not have to 
examine all electrons in the system but low energy excitations
near the Fermi level.
Low energy excitations at half filling move along 
the tubule axis because the circumference degree of freedom
(an excitation in the compactified direction) 
is frozen by a wide energy gap( $\sim 10^4 [\rm{K}] $). 
Hence this system can be described as a 1+1 dimensional system.
Furthermore in the case of metallic CNTs, the system describing
small fluctuations around the Fermi points is equivalent to the two components 
``massless" fermions in 1+1 dimensions. This allows us to analyze
the low energy excitations with sufficient accuracy 
by means of technique in quantum field theory.

Quantum mechanical behavior of the ``massless" fermions in metallic 
CNTs is governed by many kinds of interactions. 
Among others the Coulomb interaction is the most important 
interaction and drives the system into a strongly correlated system.
The purpose of this paper is to investigate how the Coulomb interaction
plays the role when the system is perturbed.
Particularly we try to answer the following 
basic question: ``What is happening when we put an external charge on a 
metallic carbon nanotube?" The external charge which we consider in this 
paper is externally fixed spinless particle. This is thought to be a charged 
impurity in a metallic nanotube~\cite{AA}.
To answer the above question we investigate the charging energy and charge
screening effect~\cite{Farajian,LT}.

This paper is composed as follows.
A model Hamiltonian for low energy excitations in metallic nanotubes 
is constructed in Sec.\ref{sec:Hamiltonian}. 
In Sec.\ref{sec:Quantization},
we rewrite the kinetic Hamiltonian in terms of current operators.
Then we examine the Hamiltonian including the long-range Coulomb interaction and 
analyze some effects of the Coulomb interaction on charging energy and 
charge screening effect in Sec.\ref{sec:Coulomb}.
In Sec.\ref{sec:gate} we consider the gate-electron interaction to show 
the effect of the long-range Coulomb interaction against an external 
perturbation.
Conclusion and discussion are given in Sec.\ref{sec:Discussion}.
An another type of the Coulomb potential is considered in App.\ref{app:2}.
We give an alternative simple derivation to obtain 
main results using the spin-charge separation in App.\ref{app:1}.

\section{The Hamiltonian}\label{sec:Hamiltonian}

Low energy excitations in metallic carbon nanotubes 
consist of four independent fermion fields. 
Their quanta are spin up($ \uparrow $) and down($ \downarrow $) electrons
in the $ K $ and $ K' $ Fermi points. We denote these fields as,
\begin{equation}
\Psi_i = \pmatrix{ \psi_{L,i} \cr \psi_{R,i} }, \ \ i \in (\uparrow,\downarrow)\times (K,K') ,
\label{eq:4fermions}
\end{equation}
where $ i $ is a label for the four fermions and these fermions are expressed by
$ \Psi_{K_\uparrow} ,\Psi_{K_\downarrow} ,\Psi_{K'_\uparrow} ,\Psi_{K'_\downarrow} $ respectively.
For example, the field $\Psi_{K_\uparrow}$ expresses the spin up electron field in the $ K $
Fermi point.
Hereafter we use $i $ as an element of the set: 
$S = \{ K_\uparrow,K_\downarrow,K'_\uparrow,K'_\downarrow \} 
\equiv (\uparrow,\downarrow)\times (K,K') $ 
as in eq.(\ref{eq:4fermions}).
Each fermion field consist of two components spinor that we name them 
$\psi_{L,i}$ and $\psi_{R,i}$.
This two components structure is due to the specific lattice structure of the 
graphite sheet~\cite{DM}.
The subscripts L and R denote left-handed component and right-handed component respectively.
The left and right are defined by the eigenvalue of the matrix 
$\gamma^5( = \sigma_3 )$ and $\sigma_3 $ is the $z$-component of the Pauli spin matrix.

Time evolution of these fields is governed by 
the complicated Hamiltonian in realistic circumstances.
For example, tubes are not strict straight line but generally bend~\cite{KM} 
and the end of a tube(cap) might mix the wave-functions of 
different Fermi points~\cite{YA}.
We restrict our attention to the most basic and important interactions:
kinetic interaction, one-dimensional long-range Coulomb interaction and gate-electron 
interaction with a simple boundary condition.
The kinetic term is 
\begin{equation}
{\cal H}_F = \sum_{i \in S}  \Psi^\dagger_i h_F \Psi_i = v_F \sum_{i \in S} \Psi^\dagger_i
\pmatrix{ P_x & 0 \cr
0 & -P_x } \Psi_i, 
\end{equation}
where $P_x= i\hbar \partial_x$ is the momentum in the 
tubule axis direction($= x \in [0:L] \equiv D$)
and $v_F $ is the Fermi velocity.
It is remarkable that the dispersion relation is linear due to the special band structure
of the $\pi$ electrons in the graphite sheet. 
Therefore we call these fermions ``massless" fermions.

The one-dimensional long-range Coulomb interaction is 
\begin{equation}
{\cal H}_C = \frac{e^2}{8\pi} \int_D \frac{J^0(x)J^0(y)}{\sqrt{|x-y|^2+d^2}} dy,
\label{eq:myCoulomb}
\end{equation}
where $e$ is the electron charge and 
$d$ in the denominator denotes the diameter of a nanotube. 
$J^0(x)$ stands for the sum of each fermion charge density 
$J^0_i(x)(\equiv \Psi_i^\dagger(x)\Psi_i(x))$,
\begin{equation}
J^0(x) = \sum_{i \in S} J^0_i(x). 
\end{equation}
A comment is in order about introducing the cutoff $d $ in the Coulomb potential 
in eq.(\ref{eq:myCoulomb}).
When we write the Coulomb interaction without the cutoff, it has 
an ultraviolet divergence in the limit of $x \to y$. To avoid the divergence, 
we need to introduce a cutoff. It is appropriate to set
it the diameter of a tube because when two electrons on a nanotube
come from opposite sides of the tubule axis direction, 
they repeal and pass each other. At the moment
they approach most, there is still a distance about the diameter of 
a nanotube to decrease the energy of such event. This is an origin of the cutoff.
Similar kind of cutoff~\cite{EG} in the Coulomb potential is derived from
integrating out the circumference degree of the Coulomb interaction. 
Both potentials have the same behavior at long distance scale $|x-y| \ge {\cal O}(d) $.
However there is a significant difference at short distance scale $|x-y| < {\cal O}(d)$.
Some detailed discussions about this potential are given in App.~\ref{app:2}.

We also include the effect of gate voltage on a nanotube.
The interaction between the gate voltage and electrons is
\begin{equation}
{\cal H}_G = e V_g \cdot J^0(x),
\end{equation}
where $V_g$ is a gate voltage which massless fermions feel.
It should be noted that the gate voltage is not usually equivalent to 
a voltage of the gate itself because of the capacitance of a substrate.
We may consider the position dependent gate voltage $V_g(x)$ 
as a model for an electrical contact or as an external perturbation to a nanotube. 
Response of the system against a local gate voltage is investigated
in Sec.\ref{sec:gate}.
By comparing the system including the long-range Coulomb interaction with 
the one excluding the interaction, 
we show how the long-range Coulomb interaction changes the response 
to the local gate voltage.

The microscopic Hamiltonian density for this system is
\begin{equation}
{\cal H} = {\cal H}_F + {\cal H}_C + {\cal H}_G.
\label{eq:Hamiltonian}
\end{equation}
We shall analyze this Hamiltonian without any approximation.
As for the screening of external particles which 
do not have a spin, we may neglect ``nonlinear interactions"\cite{EG}
such as backscattering processes because total charge sector(\ref{eq:totalc})
decouples from these interactions.

\section{Quantization of electron}\label{sec:Quantization}

We have specified the Hamiltonian density which describes the low
energy excitations in metallic nanotubes. 
In this section we quantize the massless fermion fields 
and rewrite the kinetic Hamiltonian in terms of the bosonic 
current operators~\cite{Tomonaga}. 
The Coulomb interaction and gate-electron interaction 
will be investigated in later sections.

The energy eigenfunctions of the first quantized kinetic Hamiltonian are given by
\begin{eqnarray}
&&
h_F \psi_n \pmatrix{ 1 \cr 0 } = \epsilon_n  \psi_n \pmatrix{ 1 \cr 0 },
\\
&&
h_F \psi_n \pmatrix{ 0 \cr 1 } = - \epsilon_n \psi_n \pmatrix{ 0 \cr 1 },
\\
&&
\psi_n(x) = \frac{1}{\sqrt{L}} e^{-i\frac{\epsilon_n}{\hbar v_F} x},
\end{eqnarray}
where $\epsilon_n$ is the energy eigenvalues and $L$ is the length of
a metallic nanotube. The energy eigenvalues are quantized by a boundary condition.
Here we impose the periodic boundary condition on the wave functions 
$\psi_n(x+L) = \psi_n(x)$.
This boundary condition yields the following energy spectra:
\begin{equation}
\epsilon_n = \frac{2\pi \hbar v_F}{L} n \equiv \Delta  n ,
\end{equation}
where $\Delta$ is an energy scale
which depends only on the length of a nanotube.
In order to evaluate the numerical value of this energy we use the formula for the 
Fermi velocity
\begin{equation}
v_F = \frac{3 \gamma a}{2\hbar} \sim \frac{c}{343},
\end{equation}
where $c$ is the speed of light, 
$\gamma (=2.7\ [eV]) $~\cite{Wildoer} is the overlap(hopping) integral and $a(=1.42$ [\AA])
is the nearest-neighbor distance between two carbon atoms.
For a single-wall nanotube with $L = 3\ [\rm{\mu m}]$, $\Delta $ is about $1.2\ [\rm{meV}]$.

Each massless fermion field can be decomposed into 
the left and right-handed component field as 
\begin{equation}
\Psi_i(x,t) \equiv \Psi_{L,i}(x,t) + \Psi_{R,i}(x,t),
\end{equation}
where $t$ is the time. 
The left and right-handed component fields consist of the 
energy eigenfunctions and annihilation operators of each quantum,
\begin{eqnarray}
&&\Psi_{L,i}(x,t) = \sum_{n\in Z} 
a^i_n \psi_n(x) e^{-i\frac{\epsilon_n}{\hbar}t} \pmatrix{ 1 \cr 0}, \\
&&\Psi_{R,i}(x,t) = \sum_{n\in Z} 
b^i_n \psi_n(x) e^{+i\frac{\epsilon_n}{\hbar}t} \pmatrix{ 0 \cr 1},
\end{eqnarray}
where $a^i_n,b^j_n $ are independent fermionic annihilation operator 
of left-handed component of $i$-fermion and right-handed component of $j$-fermion
satisfying the anti-commutators,
\begin{equation}
\{ a^i_n , a^{j\dagger}_m \} = \{ b^i_n , b^{j\dagger}_m \} = \delta^{ij}
\delta_{nm}.
\end{equation}
All of the other anticommutators vanish.

In 1+1 dimensions, it is possible to 
construct the fermion Fock space by acting bosonic creation operators which
are bi-linear of the fermion operators on the vacuum.
Because the Coulomb interaction consists of a product of the charge density,
it is very convenient to rewrite the kinetic term 
$H_F(= \oint {\cal H}_F)$ using the bosonic charge density operators. 
For this purpose,
it is useful to introduce the left and right currents as follows:
\begin{equation}
J^0(x) = J_L(x) + J_R(x).
\end{equation}
The left and right currents for each fermion is defined as
\begin{eqnarray}
&&J_{L,i}(x) = \Psi^\dagger_{L,i}(x) \Psi_{L,i}(x), \nonumber \\
&&J_{R,i}(x) = \Psi^\dagger_{R,i}(x) \Psi_{R,i}(x).
\end{eqnarray}
We expand these currents by the Fourier modes. We start with the left sector.
The left current can be written as follows:
\begin{eqnarray}
J_L(x) \equiv \sum_{i \in S} J_{L,i}(x)
= \sum_{n \in Z} j_L^n \frac{1}{L} e^{-i \frac{2\pi nx}{L}},
\end{eqnarray}
where each components are given by
\begin{eqnarray*}
j_L^n = \sum_{i \in S} j_{L,i}^n, \ \ 
j_{L,i}^n = \sum_{m \in Z} a_m^{i\dagger} a^i_{m+n}, \ \ j_{L,i}^0 \equiv Q_{L,i}.
\end{eqnarray*}
The Fourier components $j_{L,i}^n $(current operators) satisfies the following commutation 
relations on the fermion Fock space(current algebra).
\begin{equation}
\left[ j_{L,i}^n , (j_{L,j}^m)^\dagger \right] =  n \delta_{ij} \delta_{nm},\ \ 
(j_{L,i}^m)^\dagger = j_{L,i}^{-m}.
\label{eq:Lcurrent}
\end{equation}
We proceed to consider the right sector in the same way.
The current operators are
\begin{eqnarray}
&&
J_R(x) \equiv \sum_{i \in S} J_{R,i}(x)=  
\sum_{n \in Z} j_R^n \frac{1}{L} e^{+i \frac{2\pi nx}{L}},
\\
&&
j_R^n = \sum_i j_{R,i}^n, \ \ 
j_{R,i}^n = \sum_{m \in Z} b^{i\dagger}_{m+n} b_m^i, \ \ j_{R,i}^0 \equiv Q_{R,i}.
\nonumber
\end{eqnarray}
They satisfy the bosonic commutation relations
\begin{equation}
\left[ j_{R,i}^n , (j_{R,j}^m)^\dagger \right] = n \delta_{ij} \delta_{nm},\ \ 
(j_{R,i}^m)^\dagger = j_{R,i}^{-m}.
\label{eq:Rcurrent}
\end{equation}
The left charge $Q_{L,i} $ and right charge $Q_{R,i} $ are conserved separately because
the system is invariant under the following two independent global
transformations of fermion fields,
$\Psi_{L,i} \to e^{i\theta_{L,i}} \Psi_{L,i}, 
\Psi_{R,i} \to e^{i\theta_{R,i}} \Psi_{R,i}$.
We must rewrite the fermion Hamiltonian such that all the matrix elements are
the same as the original fermion Hamiltonian by means of current algebra.
It is well-known that the following Hamiltonian has the 
same matrix element as the original fermion Hamiltonian~\cite{Tomonaga,IM}
\begin{eqnarray}
H_F &=& \Delta \sum_{i \in S}
\Bigg[ \left( \frac{ \langle Q_i \rangle^2+
\langle Q_{5,i} \rangle^2}{4} -\frac{1}{12} \right) \nonumber \\
&& + \sum_{n > 0}  \left( (j_{L,i}^n)^\dagger j_{L,i}^n + (j_{R,i}^n)^\dagger j_{R,i}^n
\right) \Bigg] 
\end{eqnarray}
where the $U(1) $ charge $Q_i $ and the chiral charge $Q_{5,i}$ for each massless 
fermion is defined by the summention 
and subtraction between the left and right charges,
\begin{equation}
Q_i = Q_{L,i} + Q_{R,i}, \ \ \ Q_{5,i} = Q_{L,i} - Q_{R,i}.
\end{equation}
The physical meaning of the chiral charge is spatial integration of the
electric current density $ev_F(J_L(x)-J_R(x))$ in a tube.
This charge measures the left-right asymmetry of the vacuum 
and is defined by the difference of the left charge and the right charge on the vacuum.
Note that the vacuum energy in the last equation is non-vanishing, that is 
due to the finite size effect of a nanotube.
Here $\langle {\cal O} \rangle$ means the vacuum expectation value of an operator ${\cal O} $.
We have defined the second quantized vacuum by filling the negative energy modes, leaving
the positive energy modes empty.

\section{Long range Coulomb interaction}\label{sec:Coulomb}

The one-dimensional long-range Coulomb interaction in metallic nanotubes is 
given by the following potential form
\begin{equation}
H_C = \int_D {\cal H}_C dx = 
\frac{e^2}{8\pi} \int \!\!\! \int_D  \frac{J^0(x)J^0(y)}{\sqrt{|x-y|^2 + d^2}} dx dy,
\label{eq:1DmyCoulomb}
\end{equation}
where $d$ is the diameter of a nanotube and this is the traces of
the pseudo one-dimensional nature of carbon nanotubes. Total Hamiltonian 
of this system consists of the kinetic Hamiltonian and the Coulomb interaction.
The kinetic Hamiltonian is written in terms of the current operators as is shown in the 
previous section.
The Coulomb interaction can also be rewritten in terms of the current operators
because total charge density in the Coulomb interaction is 
a sum of the left and right current operators,
\begin{eqnarray}
J^0(x) = \sum_{n \in Z} \left(
(j_L^n)^\dagger + j_R^n \right)\frac{1}{L}e^{+i \frac{2\pi nx}{L}}.
\label{eq:J}
\end{eqnarray}
Hence, by replacing $J^0(x)$ by eq.(\ref{eq:J}) in the Coulomb interaction(\ref{eq:1DmyCoulomb}),
we obtain
\begin{eqnarray}
&&H_C = \frac{e^2}{8\pi L} V(0) \left( 
\sum_{i \in S} \langle Q_i \rangle  \right)^2 \nonumber \\
&& + \frac{e^2}{4\pi L} \sum_{n > 0} V(n)
\left( (j_L^n)^\dagger + j_R^n \right)
\left( j_L^n + (j_R^n)^\dagger \right).
\end{eqnarray}
We have introduced the Fourier components of the Coulomb potential,
\begin{equation}
V(n) = 2 \int_0^\pi dx \frac{\cos(2 nx)}{\sqrt{x^2 + \left( \frac{R}{L} \right)^2}},
\end{equation}
where $R $ is the circumference( $R = \pi d$ ) of a nanotube.
The Coulomb interaction is given by the sum of zero mode $V(0)$
and nonzero modes $V(n)$.
The nonzero Fourier mode($n \ne 0$) can be approximated to
the modified Bessel function $V(n) \sim 2 K_0(2n \frac{R}{L})$
quite well if $\frac{R}{L} \ll 1$.

We can include external charges in the theory by replacing the 
total charge density $J^0$ with the sum of the internal charge $J^0$
and external c-number charge density $J^0_{ex}$ in the Coulomb interaction,
\[
H_C = 
\frac{e^2}{8\pi} \int \!\!\! \int_D  \frac{(J^0(x)+J^0_{ex}(x))(J^0(y)+J^0_{ex}(y))}{\sqrt{|x-y|^2 + d^2}}dxdy.
\]
It is straight forward to derive the corresponding expression in terms of current operators
\begin{eqnarray}
H_C &=& \frac{e^2}{8\pi L} V(0) \left( 
\sum_{i \in S} \langle Q_i \rangle + j^0_{ex} \right)^2 \nonumber \\
&+&
\sum_{n > 0} \beta_n
\left( (j_L^n)^\dagger + j_R^n + (j^n_{ex})^* \right)
\left( j_L^n + (j_R^n)^\dagger + j^n_{ex} \right), \nonumber \\
\end{eqnarray}
where we introduce $\beta_n \equiv \frac{e^2}{4\pi L} V(n)$ and define the Fourier components
of the external charge as follows:
\begin{equation}
J^0_{ex}(x) = \frac{1}{L} \sum_{n \in Z} j^n_{ex} e^{-i\frac{2\pi nx}{L}}.
\end{equation}
Total charge of the external particle is given by the spatial integration of the 
c-number external charge density which is referred to $Q_{ex} $,
\begin{equation}
Q_{ex} \equiv \int_D J^0_{ex}(x) dx = j_{ex}^0.
\end{equation}

We finally obtain the long-range Coulomb interaction in terms of the current operators
which modifies the energy spectra of the original kinetic Hamiltonian that we discussed 
in the previous section.
Before we analyze the total Hamiltonian quantum mechanically,
it is important to know the order of magnitude of the long-range Coulomb interaction.
Typical energy scale of the 
Coulomb interaction($\beta_n$) of this system is strong 
as compared with the energy scale of the kinetic interaction.
This is shown by the ratio of the energy scale of the Coulomb 
interaction to the energy separation of the kinetic interaction,
\begin{equation}
\frac{\beta_n}{\Delta} = \frac{\alpha}{\pi} \frac{c}{v_F} K_0(2n\frac{R}{L}),
\label{eq:ratio}
\end{equation}
where $\alpha(\equiv \frac{e^2}{4\pi \hbar c} \sim \frac{1}{137}) $ is 
the fine-structure constant. We used the approximation formula 
$V(n) \sim 2 K_0(2n \frac{R}{L}) $ in eq.(\ref{eq:ratio}).
The value of the ratio is more than one for $n \le {\cal O}(\frac{L}{2R}) $.
Therefore this system is thought to be a strongly correlated 
system. So, we should first estimate the Coulomb energy 
scale of the physical quantities in this system.
For this purpose, we treat the Coulomb interaction roughly in the next 
subsection. Detailed analysis of the system will be given in latter subsection.

\subsection{Order estimation}

When an externally fixed particle with total charge $Q_{ex} $ is put on a metallic nanotube,
the charge density of the particle may be modeled by the delta-function,
\begin{equation}
J^0_{ex}(x) = Q_{ex} \delta(x-x_0),
\label{eq:delta}
\end{equation}
where we put the particle at $x_0 (\in [0:L]) $.
It can be thought that this external charge distribution is a model for 
a charged spinless impurity.

It is important to estimate the order of magnitude of the energy which 
we need to put a fixed external charge on a metallic nanotube.
We simply neglect the interaction between 
internal electrons and the external charge, so we set $\langle J^0(x) \rangle = 0 $.
In this case, we get 
\begin{equation}
\langle H_C \rangle = \frac{e^2}{8\pi} \int \!\!\! \int_D  
\frac{J^0_{ex}(x)J^0_{ex}(y)}{\sqrt{|x-y|^2 + d^2}}dxdy
= \frac{e^2}{8\pi} \frac{1}{d} Q_{ex}^2.
\label{eq:order}
\end{equation}
Estimation of this value for a metallic nanotube with $d = 1.4\ [{\rm nm}]$
gives $\langle H_C \rangle \sim 515 \cdot Q_{ex}^2\ [{\rm meV}] $.
As is discussed in the later section, this energy is strongly modified by 
the Coulomb interaction between internal electrons and the external charge.
The internal electrons move or rearrange in order to decrease the energy
of the system.
This means that the charge screening effect occurs.

On the other hand, 
in an experiment concerning transport on nanotubes,
an electron flows from the electric contact and 
after some period it spreads uniformly in a nanotube.
Then we have
\begin{equation}
J^0_{ex}(x) = Q_{ex} \frac{1}{L}.
\end{equation}
In this case, the Coulomb self energy is
\begin{eqnarray}
\langle H_C \rangle &=& \frac{e^2}{8\pi} \frac{1}{L} \ln 
\left[ \frac{\sqrt{1+\left( \frac{d}{L} \right)^2} + 1}{\sqrt{1+\left( \frac{d}{L} \right)^2} - 1}
\right] Q_{ex}^2 \nonumber \\
&\sim& \frac{e^2}{4\pi} \frac{1}{L} \ln \left( \frac{2L}{d} \right) Q_{ex}^2.
\label{eq:charging}
\end{eqnarray}
For a metallic nanotube with $d = 1.4\ [{\rm nm}]$ and $L = 3\ [{\rm \mu m}]$,
we obtain $\langle H_C \rangle \sim 4\cdot Q_{ex}^2\ [{\rm meV}] $.
It should be noticed that 
the eq.(\ref{eq:charging}) is the same formula that is usually used in the analysis
of the Coulomb blockade phenomena of a transport experiment of nanotubes.
This energy is called ``charging energy($E_c$)" and is not modified by the interaction 
with internal charges due to it's uniform distribution,
\begin{equation}
E_c \equiv \frac{e^2}{4\pi} \frac{1}{L} \ln \left( \frac{2L}{d} \right) .
\end{equation}
An experimental value of the charging energy for a metallic single-wall CNT with
$3.2 [{\rm \mu m}]$ in length is about $E_c = 3.8\ [{\rm meV}] $~\cite{PYD} and is very 
close to the above estimation.
This is a verification of the cutoff in the potential of the 
Coulomb interaction.

\subsection{Screening effect}

We have so far considered the long-range Coulomb interaction very roughly.
In this subsection we solve the system quantum mechanically and 
examine the charge screening effect. To investigate this effect, we
shall compute the charging energy and induced charge distribution by an external charge.
We start with one massless fermion case. This case corresponds to an approximation
that we only consider the system with one massless fermion, for example, $\Psi_{K_\uparrow}$.
Four massless fermions case(metallic nanotubes case) 
can be obtained by straight forward extension of the
one massless fermion case.

\subsubsection{one fermion case}

Total Hamiltonian is given by
\begin{widetext}
\begin{eqnarray}
H &=& \Delta \left\{ \left( \frac{ \langle Q \rangle^2+
\langle Q_5 \rangle^2}{4} -\frac{1}{12} \right)  
+ \sum_{n > 0}  \left( (j_L^n)^\dagger j_L^n + (j_R^n)^\dagger j_R^n
\right) \right\} \nonumber \\
&+& E_c \left( \langle Q \rangle + Q_{ex} \right)^2
+ \sum_{n > 0} \beta_n
\left( (j_L^n)^\dagger + j_R^n + (j^n_{ex})^* \right)
\left( j_L^n + (j_R^n)^\dagger + j^n_{ex} \right),
\end{eqnarray}
\end{widetext}
where we omit the subscript $i$ because we consider one massless fermion case. 
The kinetic term and the Coulomb term can be combined as follows:
\begin{eqnarray}
H = H_0 - \frac{\Delta}{12} + \sum_{n > 0} H_n,
\end{eqnarray}
where
\begin{eqnarray}
H_0 &=& \frac{\Delta}{4} \left[ \langle Q \rangle^2+ \langle Q_5 \rangle^2 \right]
+ E_c \left( \langle Q \rangle + Q_{ex} \right)^2,
\label{eq:H0}
\\ \nonumber 
H_n &=& \Delta \left( (j_L^n)^\dagger j_L^n + (j_R^n)^\dagger j_R^n
\right) \\
&+& \beta_n
\left( (j_L^n)^\dagger + j_R^n + (j^n_{ex})^* \right)
\left( j_L^n + (j_R^n)^\dagger + j^n_{ex} \right).\nonumber \\
\label{eq:Hn}
\end{eqnarray}
$H_0 $ is the zero mode Hamiltonian whose structure is simple and 
is already diagonalized. Hereafter we choose the vacuum with vanishing $U(1)$ charge 
($\langle Q \rangle =0 $) and the chiral charge($\langle Q_5 \rangle =0 $) 
for simplicity.
In this case we have $H_0 = E_c  Q_{ex}^2 $.
We diagonarize the Hamiltonian $H_n (n \ne 0)$
using the Bogoliubov transformation of current operators,
\begin{equation}
\pmatrix{ \tilde{j}_L^n \cr (\tilde{j}_R^n)^\dagger }
= \pmatrix{ \cosh t_n  & \sinh t_n \cr \sinh t_n & \cosh t_n }
\pmatrix{ j_L^n \cr (j_R^n)^\dagger },
\end{equation}
where
\begin{eqnarray}
&&
\sinh 2t_n = \frac{\beta_n}{E_n},\ \ 
\cosh 2t_n = \frac{1}{E_n} 
\left(  \Delta + \beta_n \right), \nonumber 
\\
&&
E_n = \Delta \sqrt{1+ \frac{2\beta_n}{\Delta}}. \nonumber
\end{eqnarray}
The Bogoliubov transformed currents($\tilde{j}_L^n,\tilde{j}_R^n $) 
satisfy the same bosonic commutation relations as the original 
current operators eq.(\ref{eq:Lcurrent}) and eq.(\ref{eq:Rcurrent}),
\begin{eqnarray}
[ \tilde{j}_L^n, (\tilde{j}_L^m)^\dagger ] = 
[ \tilde{j}_R^n, (\tilde{j}_R^m)^\dagger ] = 
n \delta_{nm}.
\end{eqnarray}
After some calculations we diagonarize $H_n $ as follows:
\begin{eqnarray}
H_n = &&
E_n \Big[ 
\left( (\tilde{j}_L^n)^\dagger + \gamma_n (j^n_{ex})^* \right)
\left( \tilde{j}_L^n + \gamma_n j^n_{ex} \right) \nonumber \\
&&
+ \left( (\tilde{j}_R^n)^\dagger + \gamma_n j^n_{ex} \right)
\left( \tilde{j}_R^n + \gamma_n (j^n_{ex})^* \right) 
+ n \Big] - \Delta n 
\nonumber \\
&&+ \beta_n\frac{\Delta^2}{E_n^2} (j_{ex}^n)^* j_{ex}^n ,
\label{eq:1fermion}
\end{eqnarray}
where $\gamma_n = \sinh 2t_n (\cosh t_n - \sinh t_n )$.
It is easy to find conditions of the vacuum $|vac_1 ;J^0_{ex} \rangle $
in the presence of the external 
charges(``$vac_1$" in the ket denotes vacuum state for `one' massless fermion.),
\begin{eqnarray}
&&
\left( \tilde{j}_L^n + \gamma_n j^n_{ex} \right) 
|vac_1;J^0_{ex} \rangle
= 0 , 
\nonumber \\
&&
\left( \tilde{j}_R^n + \gamma_n (j^n_{ex})^* \right)
|vac_1 ;J^0_{ex} \rangle = 0,  \ \ n > 0.
\label{eq:vac-condition}
\end{eqnarray}

We estimate the energy change between 
the two vacua $|vac_1; 0  \rangle $(without an external charge)
and $|vac_1 ;J^0_{ex} \rangle $(with an external charge)
\begin{eqnarray}
\delta E &=& \langle vac_1 ;J^0_{ex}|H(J^0_{ex}) |vac_1;J^0_{ex} \rangle
-\langle vac_1;0| H(0) |vac_1;0 \rangle \nonumber \\ 
&=& E_c Q_{ex}^2 + \sum_{n > 0} \beta_n \frac{\Delta^2}{E_n^2} 
(j_{ex}^n)^* j_{ex}^n \nonumber \\
&=& E_c Q_{ex}^2 + \sum_{n > 0} \frac{\beta_n}{1+ \frac{2\beta_n}{\Delta}} 
(j_{ex}^n)^* j_{ex}^n,
\label{eq:chargingEN}
\end{eqnarray}
where $ H(J^0_{ex}) $ denotes the Hamiltonian with the external charge distribution $J^0_{ex}$.
If we neglect the Coulomb interaction between internal charges and an external charge, 
as in the previous subsection, the charging energy is given by
\begin{eqnarray}
E_c Q_{ex}^2 + \sum_{n > 0} \beta_n (j_{ex}^n)^* j_{ex}^n.
\end{eqnarray}
Comparing this with the exact result(\ref{eq:chargingEN}) 
one can recognize the difference between these results. 
The effect of internal charge redistribution enters in the 
denominator of second term in eq.(\ref{eq:chargingEN}).
Therefore if the energy separation of the kinetic spectrum is huge compared with
the Coulomb interaction,
\begin{eqnarray}
\frac{\beta_n}{\Delta} \ll 1,
\end{eqnarray}
the effect of rearrangement of the internal electrons is very small 
so that the charging energy is hardly modified from the previous order estimation. 
However for metallic nanotubes, the situation is completely different as we
have mentioned at eq.(\ref{eq:ratio}).

Next, let us consider the rearrangement of internal electrons.
This rearrangement can be shown in the expectation value of the charge density 
operator:
\begin{eqnarray}
J^0(x) 
&=& \sum_{n \in Z} \left(
(j_L^n)^\dagger + j_R^n \right)\frac{1}{L}e^{+i \frac{2\pi nx}{L}} \nonumber \\
&=&  \frac{Q}{L} + \sum_{n > 0} \sqrt{\frac{\Delta}{E_n}} \left(
(\tilde{j}_L^n)^\dagger + \tilde{j}_R^n \right)\frac{1}{L}e^{+i \frac{2\pi nx}{L}}
\nonumber \\
&& \ \ \ \ + \sum_{n > 0} \sqrt{\frac{\Delta}{E_n}} \left(
\tilde{j}_L^n + (\tilde{j}_R^n)^\dagger \right)\frac{1}{L}e^{-i \frac{2\pi nx}{L}}.
\nonumber \\
\end{eqnarray}
The expectation value can be calculated using the definition of the vacuum in the 
presence of the external charge(\ref{eq:vac-condition}),
\begin{eqnarray}
&&
\langle J^0(x) \rangle_1 \equiv
\langle vac_1 ;J^0_{ex}|J^0(x) |vac_1 ;J^0_{ex} \rangle
\nonumber \\
&&=  \sum_{n > 0} \sqrt{\frac{\Delta}{E_n}} (-2\gamma_n) 
\frac{1}{L} 
\left( (j_{ex}^n)^* e^{+i \frac{2\pi nx}{L}} + j_{ex}^n e^{-i \frac{2\pi nx}{L}} \right)
\nonumber \\
&&= \sum_{n > 0}-\frac{\frac{2\beta_n}{\Delta}}{1 + \frac{2\beta_n}{\Delta}} 
\frac{1}{L} 
\left( (j_{ex}^n)^* e^{+i \frac{2\pi nx}{L}} + j_{ex}^n e^{-i \frac{2\pi nx}{L}} \right).
\nonumber \\
\end{eqnarray}
For the delta-function charge distribution(\ref{eq:delta}), we obtain
\begin{eqnarray}
&&
\langle J^0(x) \rangle_1
= Q_{ex} \sum_{n > 0} -\frac{\frac{2\beta_n}{\Delta}}{1 + \frac{2\beta_n}{\Delta}} 
\frac{2}{L}  \cos ( \frac{2\pi n}{L}(x-x_0) ).\nonumber \\
\label{eq:induced-1f}
\end{eqnarray}

\subsubsection{two fermions case}

Let us proceed to the two fermions case. Here we use 
$i = \{ 1,2 \}$ instead of, for example, $i = \{ K_\uparrow, K_\downarrow \}$ for simplicity.
Even in the two fermions case zero mode sector of the Hamiltonian is very simple and 
it gives only the charging energy $E_c Q_{ex}^2$. 
We should diagonarize the nonzero modes of the Hamiltonian.
This part is given as follows:
\begin{eqnarray}
H_n &=& \Delta \left( (j_{L,1}^n)^\dagger j_{L,1}^n + (j_{R,1}^n)^\dagger j_{R,1}^n
\right) \nonumber \\*
&+& \Delta \left( (j_{L,2}^n)^\dagger j_{L,2}^n + (j_{R,2}^n)^\dagger j_{R,2}^n
\right) \nonumber \\*
&+& \beta_n
\left( (j_{L,1}^n)^\dagger + j_{R,1}^n + (j_{L,2}^n)^\dagger + j_{R,2}^n  + (j^n_{ex})^* \right)
\nonumber \\*
&\times&
\left( j_{L,1}^n + (j_{R,1}^n)^\dagger + j_{L,2}^n + (j_{R,2}^n)^\dagger + j^n_{ex} \right).
\end{eqnarray}

First we focus on the fermion labeled by 1.
If we regard $j_{L,2}^n + (j_{R,2}^n)^\dagger + j^n_{ex} $ as $ j^n_{ex}$ in the 
previous analysis, we get the following expression instead of eq.(\ref{eq:1fermion})
\begin{eqnarray}
&&H_n =
E_n \Big[ 
\left( (\tilde{j}_{L,1}^n)^\dagger + \gamma_n \left[ (j_{L,2}^n)^\dagger + j_{R,2}^n  +  (j^n_{ex})^* \right]\right) \nonumber \\*
&& \times
\left( \tilde{j}_{L,1}^n + \gamma_n \left[j_{L,2}^n + (j_{R,2}^n)^\dagger +  j^n_{ex} \right] \right) \nonumber \\
&&
+ \left( (\tilde{j}_{R,1}^n)^\dagger + \gamma_n \left[j_{L,2}^n + (j_{R,2}^n)^\dagger + j^n_{ex} \right] \right) \nonumber \\
&& 
\times \left( \tilde{j}_{R,1}^n + \gamma_n \left[ (j_{L,2}^n)^\dagger + j_{R,2}^n  + (j^n_{ex})^* \right] \right) 
+ n \Big] - \Delta n 
\nonumber \\
&&
+\Delta \left( (j_{L,2}^n)^\dagger j_{L,2}^n + (j_{R,2}^n)^\dagger j_{R,2}^n
\right)
\nonumber \\
&&+ \beta_n \frac{\Delta^2}{E_n^2} 
\left[ (j_{L,2}^n)^\dagger + j_{R,2}^n  + (j^n_{ex})^* \right] \left[j_{L,2}^n + (j_{R,2}^n)^\dagger + j^n_{ex} \right], \nonumber \\
&&
\label{eq:2fer1}
\end{eqnarray}

The last two lines in the above equation have almost same structure of 
the one which we have analyzed in the previous subsection(\ref{eq:Hn}),
\begin{eqnarray}
&&
\Delta \left( (j_{L,2}^n)^\dagger j_{L,2}^n + (j_{R,2}^n)^\dagger j_{R,2}^n
\right)
\nonumber \\
&&+ \beta_n \frac{\Delta^2}{E_n^2} 
\left[ (j_{L,2}^n)^\dagger + j_{R,2}^n  + (j^n_{ex})^* \right] \left[j_{L,2}^n + (j_{R,2}^n)^\dagger + j^n_{ex} \right]. \nonumber \\
&&
\end{eqnarray}
However in the present case, the Coulomb interaction term is modified as follows:
\begin{eqnarray}
\beta_n \to \beta_n \frac{\Delta^2}{E_n^2}.
\end{eqnarray}

We use the same method (Bogoliubov transformation) to diagonarize 
this part of the Hamiltonian with rotation angle $s_n $ which is 
different from the previous one($t_n $).
\begin{eqnarray}
\pmatrix{ \tilde{j}_{L,2}^n \cr (\tilde{j}_{R,2}^n)^\dagger }
= \pmatrix{ \cosh s_n  & \sinh s_n \cr \sinh s_n & \cosh s_n }
\pmatrix{ j_{L,2}^n \cr (j_{R,2}^n)^\dagger },
\end{eqnarray}
where
\begin{eqnarray}
&&
\sinh 2s_n = \frac{\beta_n \frac{\Delta^2}{E_n^2}}{F_n},\ 
\cosh 2t_n = \frac{1}{F_n} 
\left(  \Delta + \beta_n \frac{\Delta^2}{E_n^2} \right), \nonumber 
\\
&&
F_n = \Delta \sqrt{1+ \frac{2\beta_n}{\Delta}\frac{\Delta^2}{E_n^2} }. \nonumber
\end{eqnarray}
This transformation gives the result which is similar to eq.(\ref{eq:1fermion}),
\begin{eqnarray}
&&
F_n \Big[ 
\left( (\tilde{j}_{L,2}^n)^\dagger + \delta_n (j^n_{ex})^* \right)
\left( \tilde{j}_{L,2}^n + \delta_n j^n_{ex} \right) \nonumber \\
&&
+ \left( (\tilde{j}_{R,2}^n)^\dagger + \delta_n j^n_{ex} \right)
\left( \tilde{j}_{R,2}^n + \delta_n (j^n_{ex})^* \right) 
+ n \Big] - \Delta n 
\nonumber \\
&&+ \beta_n \frac{\Delta^2}{E_n^2}\frac{\Delta^2}{F_n^2} (j_{ex}^n)^* j_{ex}^n ,
\label{eq:2fer2}
\end{eqnarray}
where $\delta_n = \sinh 2s_n ( \cosh s_n - \sinh s_n )$.
We define the vacuum for the two fermions case with external charge distribution.
\begin{eqnarray*}
&& \left( \tilde{j}_{L,2}^n + \delta_n j^n_{ex} \right) |vac_2 ;J^0_{ex} \rangle= 0, \\
&& \left( \tilde{j}_{R,2}^n + \delta_n (j^n_{ex})^* \right) |vac_2 ;J^0_{ex} \rangle = 0, \\
&& \left( \tilde{j}_{L,1}^n + \gamma_n \left[j_{L,2}^n + (j_{R,2}^n)^\dagger +  j^n_{ex} \right] \right) |vac_2 ;J^0_{ex} \rangle = 0, \\
&& \left( \tilde{j}_{R,1}^n + \gamma_n \left[ (j_{L,2}^n)^\dagger + j_{R,2}^n  + (j^n_{ex})^* \right] \right) |vac_2 ;J^0_{ex} \rangle = 0.
\end{eqnarray*}
From eqs.(\ref{eq:2fer1}),(\ref{eq:2fer2}) and the above definitions of the vacuum,
the energy change of the two vacuum $|vac_2;0  \rangle $ and $|vac_2 ;J^0_{ex} \rangle$
is evaluated as
\begin{eqnarray}
\delta E
&=& E_c Q_{ex}^2 + \sum_{n > 0} \beta_n \frac{\Delta^2}{E_n^2}\frac{\Delta^2}{F_n^2} 
(j_{ex}^n)^* j_{ex}^n \nonumber \\
&=& E_c Q_{ex}^2 + \sum_{n > 0} \frac{\beta_n}{1+ 2\frac{2\beta_n}{\Delta}} 
(j_{ex}^n)^* j_{ex}^n .
\end{eqnarray}
We see that the effect of `two' fermions on the charging energy 
is the factor 2 in front of the $ \frac{2\beta_n}{\Delta}$ in the denominator
of second term. 
This means that the charging energy decreases
more as compared with one fermion case.
On the other hand, 
the internal electron charge density distribution becomes
\begin{equation}
\langle J^0(x) \rangle_2
= \sum_{n > 0}-\frac{2\frac{2\beta_n}{\Delta}}{1 + 2\frac{2\beta_n}{\Delta}} 
\frac{1}{L} 
\left( (j_{ex}^n)^* e^{+i \frac{2\pi nx}{L}} + j_{ex}^n e^{-i \frac{2\pi nx}{L}} \right).
\end{equation}
It can be seen that the effect of the two fermions on the induced charge density
is also the factor 2 in front of the $\frac{2\beta_n}{\Delta} $ terms.

\subsubsection{four fermions case}

\begin{figure}
\begin{center}
\includegraphics[scale=0.6]{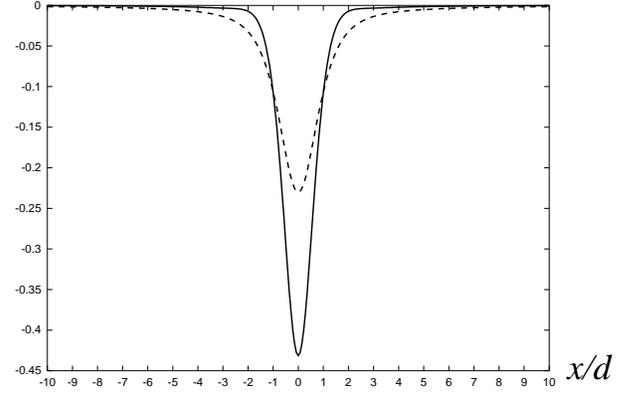}
\end{center}
\caption{Position dependence of the induced charge density in the presence of 
an external spinless point particle(\ref{eq:delta}) that has unit charge $Q_{ex}=1$.
The screening length reaches several $d$(diameter of a nanotube).
The solid line is given by the formula of the four fermions case(\ref{eq:final-delta}), 
the dashed line is for the one fermion case(\ref{eq:induced-1f}). 
This dashed line corresponds to the week coupling 
case($\frac{\alpha}{4} $) of the four fermions case.
We used $d = 1.4\ [\rm{nm}]$ and $L = 3\ [\rm{\mu m}]$ in these plots.}
\label{fig:screen-delta}
\end{figure}

Four fermions case can be analyzed by the same procedure just like two fermions case.
The calculation is straight forward but lengthy, so we shall omit them here.
Instead, we give an alternative simple derivation in App.\ref{app:1}.
The formulae for the charging energy and
induced internal charge density are 
\begin{equation}
\delta E =  E_c Q_{ex}^2 + \sum_{n > 0} \frac{\beta_n}{1+ \frac{8\beta_n}{\Delta}} 
(j_{ex}^n)^* j_{ex}^n ,
\label{eq:finalcharging} 
\end{equation}
\begin{equation} 
\langle J^0(x) \rangle_4 =
\sum_{n > 0}-\frac{\frac{8\beta_n}{\Delta}}{1 + \frac{8\beta_n}{\Delta}} 
\frac{1}{L} 
\left( (j_{ex}^n)^* e^{+i \frac{2\pi nx}{L}} + j_{ex}^n e^{-i \frac{2\pi nx}{L}} \right),
\label{eq:finalinduce}
\end{equation}
where the subscript $4 $ of the vacuum expectation value 
($\langle J^0(x)\rangle_4 $) indicates the vacuum of
the four fermions case. 
Notice that these equations can be obtained from results of
one fermion case by replacing the level spacing as follows:
\begin{eqnarray}
\Delta \to \frac{\Delta}{4}.
\label{eq:4times}
\end{eqnarray}
For the delta-function charge distribution of the external charge(\ref{eq:delta}), we obtain
\begin{eqnarray}
&&
\delta E =  \left[ E_c  + \sum_{n > 0} \frac{\beta_n}{1+ \frac{8\beta_n}{\Delta}} \right]
Q_{ex}^2, \\
&& \langle J^0(x) \rangle_4 =
Q_{ex} \sum_{n > 0}-\frac{\frac{8\beta_n}{\Delta}}{1 + \frac{8\beta_n}{\Delta}} 
\frac{2}{L}
\cos \left( \frac{2\pi n}{L}(x-x_0) \right).\nonumber \\
\label{eq:final-delta}
\end{eqnarray}
The numerical value of the above charging energy is about $85 \cdot Q_{ex}^2\ [\rm{meV}] $
for a metallic nanotube with $d = 1.4\ [\rm{nm}]$ and $L = 3\ [\rm{\mu m}]$.
This value is to be compared with the previous estimation of 
the Coulomb self-energy of the external charge 
$\langle H_C \rangle \sim 515\cdot Q_{ex}^2\ [\rm{meV}]$ in eq.(\ref{eq:order}).
This indicates that many internal electrons are influenced by the external charge
and rearranged. To confirm this we plot the expectation value of the internal charge 
density in Fig.\ref{fig:screen-delta}.

%%%%%%%%%%%%%%%%%%%%%%%%%%%%%
%%%%%%%%%%%%%%%%%%%%%%%%%%%%%

Notice that the induced charge density spreads within the range of ${\cal O}(d) $.
This fact can be recognized by the following reason.
In eq.(\ref{eq:final-delta}), we can approximately regard the summention of $n $ by integral of
$a \equiv 2n\frac{R}{L} $ if the length of a nanotube is huge compared with
the circumference($\frac{R}{L} \ll 1 $). This gives
\[
Q_{ex} \int_{0^+}^\infty -\frac{8\frac{\alpha}{\pi}
\frac{c}{v_F}K_0(a)}{1 + 8\frac{\alpha}{\pi}\frac{c}{v_F}K_0(a)} 
\frac{1}{R}
\cos \left( a\frac{x-x_0}{d} \right) da.
\]
Therefore a remaining typical length scale is the diameter of a nanotube.
Off course, the screening length depends on the physical parameter such as
the coupling constant of the Coulomb interaction($\alpha $) and the Fermi velocity.
We also plot eq.(\ref{eq:induced-1f}) of one fermion case.
The equation corresponds to the week coupling 
($\alpha \to \frac{\alpha}{4} $) of the four fermions case.
The screening length is related to a cutoff 
in the long-range Coulomb interaction, then if we take an another
type of cutoff, the screening length depends on it.

\begin{figure}
\begin{center}
\includegraphics[scale=0.6]{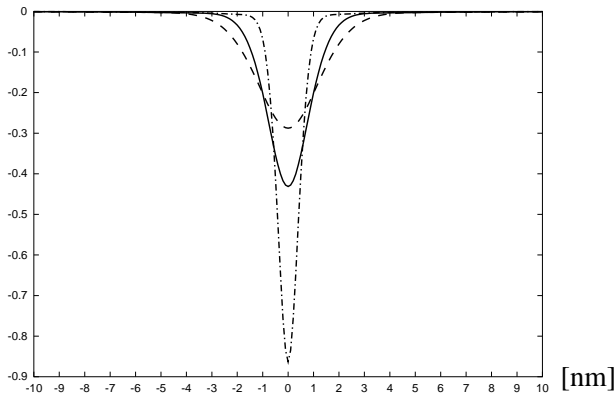}
\end{center}
\caption{Diameter dependence of the screening length.
The scale of the length is given by ${\rm nm}$.
The solid line is for $d =1.4 [\rm{nm}]$ metallic nanotube, 
the dashed line is for $d =2.1 [\rm{nm}]$ case 
and the dashed-dotted line is for $d =0.7 [\rm{nm}]$. 
We use $L = 3\ [\rm{\mu m}]$ and $Q_{ex} =1 $ in these plots.}
\label{fig:d-depened-screen-delta}
\end{figure}

Let us consider nanotubes which have different diameter. 
Fig.\ref{fig:d-depened-screen-delta} shows the induced charge 
distribution(\ref{eq:final-delta}) for nanotubes with
$d = 0.7,1.4,2.1 [{\rm nm}]$. Here we suppose that these nanotubes
have the same length $L = 3[{\rm \mu m}] $.
It should be noted that the screening length is proportional to the 
diameter of a nanotube.
%%%%%%%%%%%%%%%%%%%%%%%%%%%%%
%%%%%%%%%%%%%%%%%%%%%%%%%%%%%

\section{gate-electron interaction}\label{sec:gate}

In this section we consider the position dependent gate voltage and 
discuss the effect of the Coulomb interaction between internal electrons 
in this system.
The gate voltage is expanded in a Fourier series,
\begin{eqnarray}
V_g(x) = g \sum_{n \in Z} v_g^n e^{-i\frac{2\pi nx}{L}},
\end{eqnarray}
where $g $ has the dimension of voltage.
In this case the gate-electron interaction consist of
the zero mode and nonzero modes,
\begin{eqnarray}
H_G = eg v^0_g Q &+& \sum_{i \in S} \sum_{n >0} 
eg \left[ v_g^n (j_{L,i}^n)^\dagger + (v_g^n)^* j_{L,i}^n \right] \nonumber \\*
&&+eg \left[ v_g^n j_{R,i}^n + (v_g^n)^* (j_{R,i}^n)^\dagger \right].
\end{eqnarray}

\begin{figure}
\begin{center}
\includegraphics[scale=0.6]{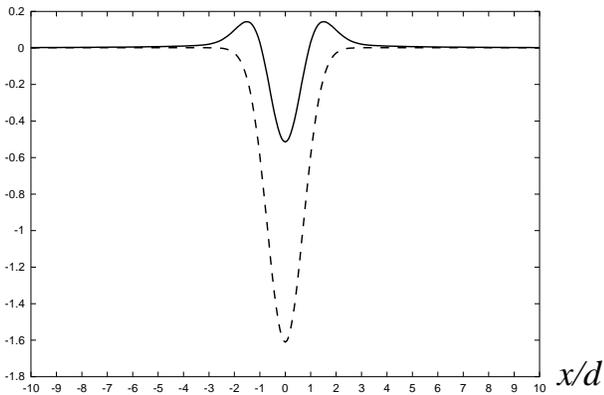}
\end{center}
\caption{Position dependence of the induced charge density 
by the local gate voltage(\ref{eq:gate-pot}).
The solid line is given by the formula(\ref{eq:induced-gate}) and 
the dashed line is given by the same formula however 
without the Coulomb interaction($\beta_n = 0$).
We set $\frac{2eg}{\Delta}=1 $ for a metallic nanotube with $d = 1.4 [{\rm nm}]$ and
$L = 3 [{\rm \mu m}] $.}
\label{fig:screen-gate}
\end{figure}

Let us concentrate on the nonzero modes of one fermion case
without an external charge.
The Hamiltonian for the nonzero modes is rewritten as
\begin{eqnarray}
H_n = &&
E_n \Big[ 
\left( (\tilde{j}_L^n)^\dagger + \Gamma_n (v_g^n)^* \right)
\left( \tilde{j}_L^n + \Gamma_n v_g^n \right) \nonumber \\
&&
+ \left( (\tilde{j}_R^n)^\dagger + \Gamma_n v_g^n \right)
\left( \tilde{j}_R^n + \Gamma_n (v_g^n)^* \right) + n \Big] -\Delta n 
\nonumber \\
&& - 2 E_n \Gamma_n^2 (v_g^n)^* v_g^n,
\end{eqnarray}
where
\begin{eqnarray}
\Gamma_n = \frac{eg}{E_n} \sqrt{\frac{\Delta}{E_n}}.
\end{eqnarray}
The above Hamiltonian gives us definitions of the vacuum,
\begin{eqnarray}
&&
\left( \tilde{j}_L^n + \Gamma_n v_g^n \right) 
|vac_1;V_g \rangle
= 0 , 
\nonumber \\
&&
\left( \tilde{j}_R^n + \Gamma_n (v_g^n)^* \right)
|vac_1;V_g \rangle = 0,  \ \ n > 0.
\label{eq:}
\end{eqnarray}
With these definitions we can compute the expectation 
value of the charge density operator.
As a simple example, 
we use a local gate potential which has the Gaussian form
\begin{eqnarray}
V_g(x) \sim g \sqrt{\frac{k}{\pi}} e^{-k(x-x_0)^2},\ \ 
v_g^n \sim e^{-\frac{\pi^2 n^2}{k L^2}} e^{i \frac{2\pi n x_0}{L} },
\label{eq:gate-pot} 
\end{eqnarray}
where $k $ decides the size of the local gate potential.
Here we set $k = 1/d^2$.
This gives the induced charge density for four fermions case,
\begin{equation}
\langle J^0(x) \rangle_4 = - \sum_{n >0}
\frac{4\frac{2eg}{\Delta}}{1+ 4\frac{2\beta_n}{\Delta}}
\frac{2}{L} e^{-\left( \frac{R}{L}\right)^2 n^2} \cos \left( \frac{2\pi n}{L}(x-x_0) \right).
\label{eq:induced-gate}
\end{equation}
In order to see the effect of the long-range
Coulomb interaction on the induced charge density,
we plot this function with the Coulomb interaction 
and without it($\beta_n = 0 $) in Fig.\ref{fig:screen-gate}.
The plots in Fig.\ref{fig:screen-gate} show that
the long-range Coulomb interaction between internal electrons significantly
changes the response of the system to the external perturbation. 
The effect of the Coulomb interaction on the induced charge density
appears in the denominator of the above result.
It should be noted that finite width of the local gate voltage makes the 
effective range of $n $ in the summention small and high frequency modes 
of the potential ineffective.

%%%%%%%%%%%%%%%%%%%%%%%%%%%%%
%%%%%%%%%%%%%%%%%%%%%%%%%%%%%

\section{discussion and comments}\label{sec:Discussion}

In this paper we have analyzed the charge screening effect 
in metallic carbon nanotubes. 
The significance of our present work is put as follows.

We modeled the Hamiltonian(\ref{eq:Hamiltonian}) describing 
the low energy excitations in metallic nanotubes
and solved the system in the presence of the external charge.
It was found that when we put an external particle
on a metallic nanotube, the electric charge of the particle is screened by internal electrons
due to the long-range Coulomb interaction between the particle and internal electrons.
The Coulomb interaction is strong as compared with 
the energy scale of the kinetic Hamiltonian(\ref{eq:ratio}).
This fact makes the quantum mechanical screening complete.

The screening length is given by about the diameter of a nanotube
in regard to the long-range Coulomb potential(\ref{eq:1DmyCoulomb}).
However the length depends on the cutoff in the potential of the 
Coulomb interaction(see App.\ref{app:2}),
therefore it is important to understand the short distance behavior of the 
potential to find more accurate value of the screening length. 
This is a nontrivial problem because the short distance corresponds 
to the high energy region, hence we necessarily examine other bands
in addition to the linear bands.
Anyway the effective range of screening(about the diameter of a nanotube or less than that)
is very small compared with the length of a nanotube,
the end of a nanotube(cap) can be thought to be ineffective to the screening phenomena.

The formula for the induced charge density in the presence of
a point particle is given by eq.(\ref{eq:final-delta}). 
The summention of $n$ in this equation converges due to eq.(\ref{eq:ratio}).
Because of that an extra cutoff of $n$ is not necessary~\cite{DM}.
In substance the summention converges up to $n \sim {\cal O}(\frac{L}{d})$.
However, 
the high frequency modes$(n > {\cal O}(\frac{L}{d}))$ of the long-range Coulomb interaction might 
be influenced by the other bands which do not belong to the `massless' dispersion
bands. 
We would like to make a quantitative analysis of these massive bands
in a future report.

%%%%%%%%%%%%%%%%%%%%%%%
\begin{acknowledgments}
The author would like to thank T. Ando and A. Farajian for fruitful discussion.
This work is supported by a fellowship of the Japan 
Society of the Promotion of Science.
\end{acknowledgments}

\appendix
\section{}\label{app:2}

\begin{figure}
\begin{center}
\includegraphics[scale=0.6]{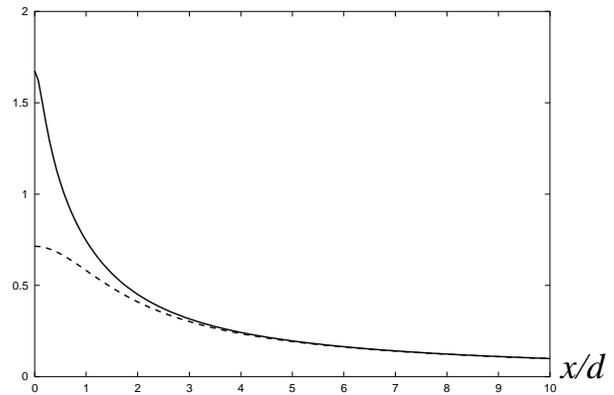}
\end{center}
\caption{Position dependence of the Coulomb potentials.
The solid line is given by the formula (\ref{eq:coulomb2}) and 
the dashed line is the potential that we used in the text.
Here we take $d = 1.4 [{\rm nm}]$ and $a_z = 1.4$ [\AA].}
\label{fig:potential}
\end{figure}

We may use an another kind of the long-range Coulomb potential which is derived from
integrating out the circumference degree of the Coulomb interaction 
with a cutoff $a_z$~\cite{EG}. The interaction is given by,
\begin{eqnarray}
H_C = \frac{e^2}{8\pi} \int \!\!\! \int_D  J^0(x)V(x-y)J^0(y) dx dy,
\end{eqnarray}
with the potential,
\begin{eqnarray}
V(x) = \frac{1}{\sqrt{|x|^2 + d^2 + a_z^2}}
\frac{2}{\pi} K \left( \frac{d}{\sqrt{|x|^2 + d^2 + a_z^2}} \right).
\label{eq:coulomb2}
\end{eqnarray}
$K(z) $ is the complete elliptic integral of the first kind
and the cutoff $a_z(\sim a)$ denotes the average distance between a $2p_z $ electron
and the nucleus.
It should be noted that a new length scale $a_z$ in addition to $L$(length of a nanotube)
and $d$(diameter) is coming out.
This scale modifies the behavior of the potential at short distance($x \to 0 $)
as is shown in Fig.\ref{fig:potential}. 
%%%%%%%%%%%%%%%%%%%%%%%%%%%%%
%%%%%%%%%%%%%%%%%%%%%%%%%%%%%

\begin{figure}
\begin{center}
\includegraphics[scale=0.6]{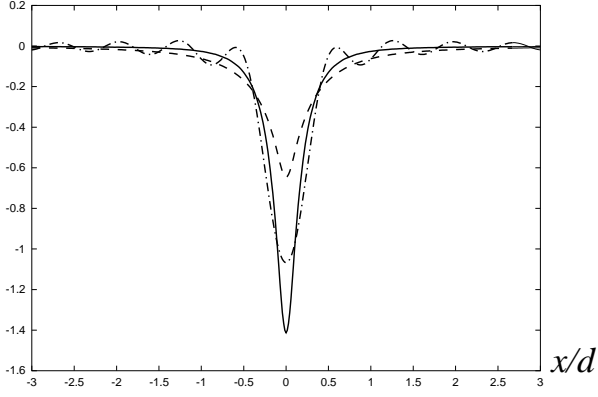}
\end{center}
\caption{Position dependence of the induced charge density
by an external spinless point particle(\ref{eq:delta}) with $Q_{ex}=1$.
The solid line is given by the formula for the four fermions case(\ref{eq:final-delta}), 
the dashed line is for the one fermion case(\ref{eq:1fermion}). 
The dotted-dashed line shows the function in eq.(\ref{eq:cutoff}) that is
regularized by the function(\ref{eq:reg}) with $n^* = 3000 \sim {\cal O}(\frac{L}{d})$.
We use the Fourier components$(\ref{app:Fourier})$ and take $a_z = 1.4(\sim a)$[\AA],
$d = 1.4\ [\rm{nm}]$ and $L = 3\ [\rm{\mu m}]$ in these plots.}
\label{fig:induced-egger}
\end{figure}

Hence Fourier components of the potential,
\begin{eqnarray}
V(n) = \frac{2}{\pi} \int^{\frac{\pi}{2}}_0 2 K_0 \left( 2n \frac{R}{L}\sqrt{\sin^2 x + \left(\frac{a_z}{d}\right)^2 } \right)dx
\label{app:Fourier}
\end{eqnarray}
are different from the previous one $V(n) \sim 2K_0( 2n\frac{R}{L}) $.
We plot the induced charge density using eq.(\ref{app:Fourier}) in Fig.\ref{fig:induced-egger}.
It can be seen that the screening length is about one half the length of the diameter.

It is important to note that the effective range of $n$ summention in 
the formula of the induced charge density becomes large 
compared with the Coulomb interaction in the text.
Within this plot, the summention converges up to $n \sim {\cal O}(\frac{L}{\pi a_z}) $.
So, it is not clear if the screening effect can be recognized in the framework 
of the ``low energy" excitations.
In order to answer such question, we introduce a regulator to examine 
if the high frequency modes of the potential give a significant contribution 
to the final result~\cite{DM}. Here we take a simple regulator,
\begin{eqnarray}
{\rm reg}(n) = \frac{1}{\exp[ n-n^*] + 1},
\label{eq:reg}
\end{eqnarray}
where $n^* \sim {\cal O}(\frac{L}{d})$.
We define the following induced charge density,
\begin{equation}
\langle J^0(x) \rangle_4^{\rm reg} \equiv
 \sum_{n > 0}-  \frac{{\rm reg}(n) \frac{8\beta_n}{\Delta}}{1 + \frac{8\beta_n}{\Delta}} 
\frac{2}{L}
\cos \left( \frac{2\pi n}{L}(x-x_0) \right).
\label{eq:cutoff}
\end{equation}
This function is also shown in Fig.\ref{fig:induced-egger}.
We see from this figure that the induced charge density oscillates.
This is due to the regulator and does not have any physical meaning. 
What needs to be emphasized at this point is that 
the screening effect arises from the contribution of the low energy region $n < n^* $.
Thus it is concluded that the charge screening can be analyzed in the
low energy physics.

\begin{figure}
\begin{center}
\includegraphics[scale=0.6]{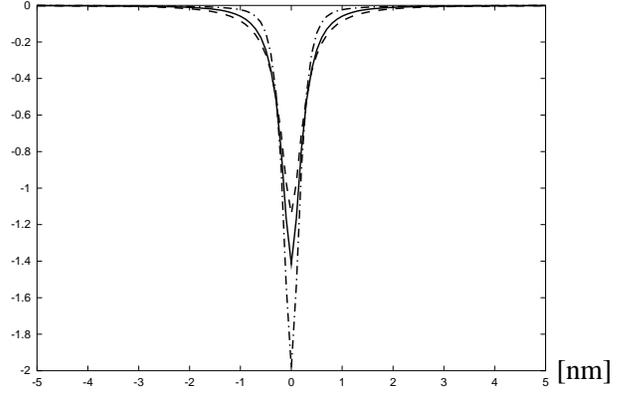}
\end{center}
\caption{Diameter dependence of the screening length.
The scale of the length is given by ${\rm nm}$.
The solid line is for $d =1.4 [\rm{nm}]$ metallic nanotube, 
the dashed line is for $d =2.1 [\rm{nm}]$ case 
and the dashed-dotted line is for $d =0.7 [\rm{nm}]$. 
We use the Fourier components$(\ref{app:Fourier})$ and take $a_z = 1.4(\sim a)$[\AA] and
$L = 1\ [\rm{\mu m}]$ in these plots.}
\label{fig:d-depened-egger-screen-delta}
\end{figure}

Let us consider nanotubes which have different diameter. 
Fig.\ref{fig:d-depened-egger-screen-delta} shows the induced charge 
distribution for nanotubes with
$d = 0.7,1.4,2.1 [{\rm nm}]$ by means of the 
Coulomb potential (\ref{eq:coulomb2}). 
Here we suppose that these nanotubes
have the same length $L = 1[{\rm \mu m}] $.
It should be noted that the screening length is proportional to the 
diameter of a nanotube. However as compared with the previous potential
case (Fig.\ref{fig:d-depened-screen-delta}) the diameter dependence is rather week.

%%%%%%%%%%%%%%%%%%%%%%%%%%%%%
%%%%%%%%%%%%%%%%%%%%%%%%%%%%%

%%%%%%%%%%%%%%%%%%%%%%%%%%%%%
%%%%%%%%%%%%%%%%%%%%%%%%%%%%%

\section{}\label{app:1}

Here we derive the final results eq.(\ref{eq:finalcharging}) and eq.(\ref{eq:finalinduce}),
using the operator decomposition into spin and charge.
First we consider the density operator for $K $ Fermi point.
We define the following operators for the $K $ point
\begin{eqnarray}
&& J_{L,C_K} = J_{L,K_\uparrow} + J_{L,K_\downarrow},
\nonumber \\
&& J_{L,S_K} = J_{L,K_\uparrow} - J_{L,K_\downarrow},
\end{eqnarray}
where we omit spatial dependence of these operator for simplicity and
$J_{L,C_K},J_{L,S_K}$ express the charge and spin operators of left handed fermions 
in the $K $ Fermi point.
Similarly for the right-handed sector at $K $ Fermi point, we define
\begin{eqnarray}
&& J_{R,C_K} = J_{R,K_\uparrow} + J_{R,K_\downarrow}, 
\nonumber \\
&& J_{R,S_K} = J_{R,K_\uparrow} - J_{R,K_\downarrow}.
\end{eqnarray}
The commutation relations between these operators can be calculated 
by using the commutation relation of original operators($J_{L,i},J_{R,i} $).
For example, we obtain the following commutation relation
\begin{eqnarray}
[ J_{L,C_K}(x), J_{L,C_K}(y) ] = 2 \frac{i}{2\pi} \partial_x \delta (x-y).
\end{eqnarray}
For $K' $ Fermi point, similarly we define
\begin{eqnarray*}
&& J_{L,C_{K'}} = J_{L,K'_\uparrow} + J_{L,K'_\downarrow},\
J_{L,S_{K'}} = J_{L,K'_\uparrow} - J_{L,K'_\downarrow},\\
&& J_{R,C_{K'}} = J_{R,K'_\uparrow} + J_{R,K'_\downarrow},\
J_{R,S_{K'}} = J_{R,K'_\uparrow} - J_{R,K'_\downarrow}.
\end{eqnarray*}
Using above definitions of the density operators,
we define the symmetric and antisymmetric combination 
concerning two Fermi points using the above 
spin and charge operators for left-handed sector as follows:
\begin{eqnarray*}
&& J_{L,C_+} = J_{L,C_K} + J_{L,C_{K'}},\
J_{L,C_-} = J_{L,C_K} - J_{L,C_{K'}}, \\
&& J_{L,S_+} = J_{L,S_K} + J_{L,S_{K'}},\
J_{L,S_-} = J_{L,S_K} - J_{L,S_{K'}}.
\end{eqnarray*}
Similarly for right-handed sector,
\begin{eqnarray*}
&& J_{R,C_+} = J_{R,C_K} + J_{R,C_{K'}},\
J_{R,C_-} = J_{R,C_K} - J_{R,C_{K'}}, \\
&& J_{R,S_+} = J_{R,S_K} + J_{R,S_{K'}},\
J_{R,S_-} = J_{R,S_K} - J_{R,S_{K'}}. 
\end{eqnarray*}
We get the following kinetic Hamiltonian in term of these new density operators
\begin{eqnarray}
H_F &&=
\Delta \left[  \frac{L}{8} \int_D : J_{L,C_+}(x)^2 + J_{R,C_+}(x)^2 : dx - \frac{1}{12}  \right] 
\nonumber \\*
&&+ 
\Delta \left[  \frac{L}{8} \int_D : J_{L,C_-}(x)^2 + J_{R,C_-}(x)^2 : dx - \frac{1}{12}  \right]
\nonumber \\*
&&+ 
\Delta \left[  \frac{L}{8} \int_D : J_{L,S_+}(x)^2 + J_{R,S_+}(x)^2 : dx - \frac{1}{12}  \right]
\nonumber \\*
&&+ 
\Delta \left[  \frac{L}{8} \int_D : J_{L,S_-}(x)^2 + J_{R,S_-}(x)^2 : dx - \frac{1}{12}  \right].
\nonumber 
\end{eqnarray}
Let us define the following operator and this is actually equivalent to the 
charge density operator,
\begin{eqnarray}
&& J_{C_+} = J_{L,C_+} + J_{R,C_+}, \\
&& J^0(x) = J_{C_+}(x).
\end{eqnarray}
Therefore we find that the total charge sector is decouple from other operators
and this is famous spin and charge separation in one dimensional systems.
The Coulomb interaction is written only by the total charge density. Hence 
we combine the kinetic term and the long-range Coulomb term 
and define $C_+ $ sector as follows:
\begin{eqnarray}
&&H_{C_+} \equiv
\Delta \left[  \frac{L}{8} \int_D : J_{L,C_+}(x)^2 + J_{R,C_+}(x)^2 : dx - \frac{1}{12}  \right] 
\nonumber \\
&&+ \frac{e^2}{8\pi} \int \!\!\! \int_D  \frac{(J_{C_+}(x)+J^0_{ex}(x))(J_{C_+}(y)+J^0_{ex}(y))}{\sqrt{|x-y|^2 + d^2}}dxdy.\nonumber \\
\label{eq:totalc}
\end{eqnarray}
This total Hamiltonian is very similar to the one fermion case which we have analyzed.
The current operators is defined by the same way in the previous section
\begin{equation}
J_{C_+}(x) = \sum_{n \in Z} \left(
(j_{L,C_+}^n)^\dagger + j_{R,C_+}^n \right)\frac{1}{L}e^{+i \frac{2\pi nx}{L}}.
\end{equation}
Notice that the commutation relation is modified
\begin{eqnarray}
[ J_{L,C_+}(x), J_{L,C_+}(y) ] = 4\frac{i}{2\pi} \partial_x \delta(x-y),
\end{eqnarray}
and we obtain the current algebra
\begin{eqnarray}
&& \left[ j_{L,C_+}^n , (j_{L,C_+}^m)^\dagger \right] =  4n \delta_{nm}, \\
&& \left[ j_{R,C_+}^n , (j_{R,C_+}^m)^\dagger \right] =  4n \delta_{nm}.
\end{eqnarray}
Analysis of the above Hamiltonian can be done like one fermion case. 
we decompose it into zero mode and nonzero modes,
\begin{eqnarray}
H_{C_+} = H_0 - \frac{\Delta}{12} + \sum_{n > 0} H_n,
\end{eqnarray}
where
\begin{eqnarray}
H_0 &=& \frac{\Delta}{16} \left[ \langle Q \rangle^2+ \langle Q_5 \rangle^2 \right]
+ E_c \left( \langle Q \rangle + Q_{ex} \right)^2 \\
H_n &=& \frac{\Delta}{4} \left( (j_{L,C_+}^n)^\dagger j_{L,C_+}^n + (j_{R,C_+}^n)^\dagger j_{R,C_+}^n
\right) \nonumber \\
&+& \beta_n
\left( (j_{L,C_+}^n)^\dagger + j_{R,C_+}^n + (j^n_{ex})^* \right)
\nonumber \\
&& \times
\left( j_{L,C_+}^n + (j_{R,C_+}^n)^\dagger + j^n_{ex} \right),
\end{eqnarray}
We can get this Hamiltonian by the following replacement
in the one fermion Hamiltonian(\ref{eq:H0},\ref{eq:Hn}), see alos eq.(\ref{eq:4times}),
\begin{eqnarray}
\Delta \to \frac{\Delta}{4}.
\end{eqnarray}
Therefore formulae of the charging energy and induced charge
density for four fermions case can be obtained 
by the above replacement in the equations which was obtained 
in the analysis of the one fermion case.

%%%%%%%%%%%%%%%%%%%%%%%%

%%%%%%%%%%%%%%%%%%%%%%%%%

\end{document}